\documentclass[12pt]{article}
\usepackage{amssymb,latexsym}
\usepackage[cmex10]{amsmath}
\title{On Sequences with a Perfect Linear Complexity Profile}
\author{Graham H. Norton\footnote{
School of Mathematics and Physics, University of Queensland, Brisbane, Queensland 4072, Australia (e-mail: ghn@maths.uq.edu.au).}}
\newtheorem{theorem}{\bf Theorem}[section] 
\newtheorem{corollary}[theorem]{\bf Corollary} 
\newtheorem{proposition}[theorem]{\bf Proposition} \newtheorem{definition}[theorem]{\bf Definition} \newtheorem{lemma}[theorem]{\bf Lemma} \newtheorem{example}[theorem]{\bf Example}  
\newtheorem{algorithm}[theorem]{\bf Algorithm} 
\newtheorem{remark}[theorem]{\bf  Remark} \newtheorem{remarks}[theorem]{\bf  Remarks} 
\newenvironment{proof}{{\bf Proof.}}{\hspace*{\fill}$\blacksquare$\par\vspace{4mm}} 

\def \bt{ \begin{theorem} }
\def \et{ \end  {theorem} }
\def \bl{ \begin{lemma} }
\def \el{ \end  {lemma} }
\def \bp{ \begin{proposition} }
\def \ep{ \end  {proposition} }
\def \bc{ \begin{corollary} }
\def \ec{ \end  {corollary} }
\def \bd{ \begin{definition} }
\def \ed{ \end  {definition} }
\def \bdp{ \begin{definitionprop} }
\def \edp{ \end  {definitionprop} }
\def \bdt{ \begin{definitiontheorem} }
\def \edt{ \end  {definitiontheorem} }
\def \bpr{ \begin{proof} }
\def \epr{ \end  {proof} }
\def \ba{ \begin{algorithm} }
\def \ea{ \end{algorithm} }
\def \be{ \begin{example} }
\def \eex{ \end{example} }
\def \bes{ \begin{examples} }
\def \eexs{ \end{examples} }
\def \br{ \begin{remark} }
\def \er{ \end{remark} }
\def \brs{ \begin{remarks} }
\def \ers{ \end{remarks} }
\def \bpb{ \begin{problem} }
\def \epb{ \end{problem} }

\newcommand{\Ann} {\mathrm{Ann}}
\newcommand{\Min} {\mathrm{MP}}

\newcommand{\ol} {\overline}

\newcommand{\ul} {\underline}
\newcommand{\ee} {\mathrm{e}}

\newcommand{\hgt} {\mathrm{ht}}

\newcommand{\pp} {\mathrm{p}}
\newcommand{\ra} {\rightarrow}
\newcommand{\Ra} {\Rightarrow}
\newcommand{\vv} {\mathrm{v}}
\newcommand{\LC} {\mathrm{L}}

\newcommand{\F}{\mathrm{F}}
\newcommand{\N}{\mathrm{N}}
\newcommand{\Z}{\mathrm{Z}}

\newcommand{\tfae} {The following are equivalent:}
\begin{document}
\maketitle
\begin{abstract}
\noindent We derive B\'ezout identities for the minimal polynomials of a finite sequence and use them to prove a theorem of Wang and Massey on binary sequences with a perfect linear complexity profile. We give a new proof of Rueppel's conjecture and simplify Dai's original proof. We obtain short proofs of results of Niederreiter relating  the linear complexity of a sequence $s$ and $K(s)$, which was  defined using continued fractions. We give an upper bound for the sum of the linear complexities of any sequence. This bound is tight for sequences with a perfect linear complexity profile and we apply it to characterise these sequences in two new ways. 
\end{abstract}

\section{Introduction}
In \cite{N2010c} we showed how to obtain a minimal polynomial (MP) of a finite sequence recursively. We begin with a compact matrix reformulation of this: multiplying an updating matrix and the current  'MP matrix' gives the new one. The product rule for determinants gives a quick proof of B\'ezout identities for MP's (which was proved from first principles in \cite {N2010c}). Next we give several basic characterisations of sequences with a perfect linear complexity profile (PLCP) in terms of MP's. 

Section \ref{WM} applies the B\'ezout identities to give a new proof of a theorem of Wang and Massey characterising  binary sequences with a PLCP. 

 We give a new proof of Rueppel's conjecture and simplify the proof of \cite{Dai}. This result is that  the linear complexity (LC)  of the first $n$ terms of the binary sequence  $r=(1,1,0^1,1,0^3,1,0^7,1,...)$ is $\lfloor \frac{n+1}{2}\rfloor$.  We will see that the updating matrices of $r$ are either a constant matrix $U$ or an identity matrix. (When applying the Euclidean algorithm to $r$ of length a power of two, the successive remainders can be obtained from the matrices $M,U,\ldots,U,M$, \cite[Lemma 3]{Dai}; we obtain $M$ as the MP matrix of the first two terms of the sequence.) Our approach is to work directly with a family of binary polynomials $\{\gamma^{(i)}:\ i\geq 0\}$ rather than with the roots of $Y^2+xY+1$ in some algebraic extension of $\F_ 2(x)[Y]$, as in \cite{Dai}. (In fact,  the polynomial $\theta_n$  of \cite{Dai} equals $x\gamma^{(n)}$ and  the requisite lemmas of \cite{Dai} can be easily deduced from ours.) This means that Rueppel's conjecture can  be proved more simply: use the Euclidean algorithm and properties of the $\{\gamma^{(i)}\}$.

Section \ref{hgt} studies a quantity $K(s)$ which was defined using the continued fraction expansion of the generating function of $s$ in \cite{Nied86a}.  In \cite{Nied86b}, the author shows that $K(s)$ equals the supremum of the quantities $\ee_n$ which figure prominently in our main theorem (giving MP's recursively). We take their supremum as our starting point, defining the {\em height  of $s$} to be $\hgt(s)=\sup\{\ee_n\}$. We give short proofs of inequalities which $\hgt(s)$ satisfies (originally proved  for $K(s)$ using continued fractions) as well a characterisation of PLCP sequences in terms of $K(s)$ which appeared in \cite{Nied86a}. In this way, we can deduce results of \cite{Nied86a}, \cite {Nied86b} as corollaries.
 
 We conclude with an upper bound for the sum of the LC's of a sequence. This bound is tight for PLCP sequences and leads  to two new properties of these sequences which are equivalent to Rueppel's original definition.  
 
We thank an anonymous referee for  Theorem \ref{stable} which improves an earlier approach.
Some of the results of this paper were presented  in May 2010 at Projet Secret, INRIA, Rocquencourt, France. We would also like to thank the project members for their interest and hospitality, and Nicolas Sendrier for a useful question. 
\section{Preliminaries}
\subsection{Notation}
The letter $n$ always denotes $n$ a strictly positive integer, $\N=\{1,2,\ldots\}$, $\Z=\{0,\pm 1,\ldots\}$, $\F$ denotes a field, $\F_q$ denotes a finite field of order $q$ and $D$ is a commutative integral domain with $1\neq 0$.  For any set $S$ containing 0, $S^\times=S\setminus\{0\}$.  

As usual, $D[[x^{-1},x]$ is the domain of Laurent series in $x^{-1}$,  $D[x^{-1},x]$ is the domain of Laurent polynomials and $\vv:D[[x^{-1},x]\ra \Z\cup \{-\infty\}$ is the exponential valuation: $\vv(0)=-\infty$ and if $g\neq 0$,  $\vv(g)=\max\{i: g_i\neq 0\}$, so that $\vv$ coincides with $\deg$ on $D[x]$. 

It is elementary that $\vv(gg')=\vv(g)+\vv(g')$, $\vv(g+g')\leq \max\{\vv(g),\vv(g')\}$ and $\vv(g+g')= \max\{\vv(g),\vv(g')\}$ if $\vv(g)\neq \vv(g)'$. We also use $\vv$ denote its restriction to  $D[x^{-1},x]$. 
 
\subsection{Sequences}
An infinite sequence is a function $s:\N \rightarrow$ $D$. The set of infinite sequences over $D$  clearly forms an abelian group. We can regard it as a submodule of a natural $D[x]$-module as in   \cite[Section 2]{N95b} as follows. 
 We begin with  $D[[x^{-1}]$ as standard $D[[x^{-1}]$-module i.e. acting on itself via multiplication. This also makes $D[[x^{-1}]$ into a $D[x]$-module. Let $\ul{s}=\sum_{j\geq 1}s_jx^{-j}$. Then for $f\in D[x]$, put
$$f\circ \ul{s}=\sum_{j\geq 1}(f\cdot \ul{s})_{-j}\ x^{-j}.$$
 One checks that $\circ$ makes $x^{-1}D[[x^{-1}]]$ into a $D[x]$-module.
\begin{definition} \label{lrs} An infinite sequence $s$ satisfies a {\em linear recurrence relation}  if it is a torsion element i.e. if $\Ann(s)=\{f\in D[x]:\ f\circ \ul{s}=0\}\neq \{0\}$. In other words,  if for some $f\in D[x]^\times$, $(f\cdot \ul{s})_{d-j}=0$ for $d-j\leq -1$ where $d=\deg(f)$ i.e.
$$ f_0s_{j-d}+\cdots+f_ds_ {j}=0\mbox{ for } d+1\leq j.
$$
\end{definition}
 When $f_d=1$, we can write 
$s_{j}=-(f_0s_{ j-d}+\cdots+ f_{ d-1}s_{ j-1})$ for $j\geq d+1$ and $s$ is a {\em linear recurring sequence}. 

 A finite sequence is $s=(s_1,\ldots,s_n)\in D^n$ and
$$\ul{s}=s_nx^{-n}+\cdots +s_1x^{-1}.$$   For $1\leq i\leq n$, we write $s^{(i)}$ for $(s_1,\ldots,s_i)$.  
In the following definition, the multiplication of $f\in D[x]$ and $\ul{s}\in x^{-1}D[x^{-1}]$ is in  $D[x,x^{-1}]$.
\begin{definition} [Annihilator] \label{anndefn}(\cite[Definition 2.7, Proposition 2.8]{N95b}) We say that $f\in D[x]$  is an annihilator (or a characteristic polynomial) of $s\in D^n$ if $f=0$ or $(f\cdot \ul{s})_{d-j}=0$ for $d-n\leq d-j\leq -1$ i.e. 
$$f_0s_{j-d}+\cdots+f_ds_ {j}=0\mbox{ for } d+1\leq j\leq n$$
where $d=\deg(f)\geq 0$, written $f\in \Ann(s)$.
\end{definition}

The following definition is a functional version of \cite[Definition 2.10]{N95b} $\Delta: D[x]^\times\times D^{n+1}\ra D$ is given by 
$$\Delta(f,s)=(f\cdot\ul{s})_{d-n-1}=\sum_{k=0}^{d}f_k \ s_{n+1-d+k}$$
where $d=\deg(f)$.  If $s\in D^{n+1}$ is understood, we write $\Delta_{n+1}(f)$ for  $\Delta(f,s)$;  if $f$ is also understood, we simply write $\Delta_{n+1}$.  Clearly $\Delta_{n+1}(1,s)=s_{n+1}$ and $f\in\Ann(s)$ if and only if $f\in\Ann(s^{(n)})$ and $\Delta_{n+1}=0$.
\section{MP-matrices of a sequence}\label{MP}
The principal result in this section is Theorem \ref{bimrt}. \subsection{Minimal Polynomial} Any polynomial of degree at least $n$ annihilates $s\in D^n$ vacuously, so $\Ann(s)\neq (0)$ and the following definition makes sense.
\begin{definition}[Minimal Polynomial (MP)] \label{mindefn} (\cite[Defn. 3.1]{N95b}) We say that $f\in \Ann(s)$ is a minimal polynomial (MP) of $s\in D^n$ if  $$\deg(f)=\min\{\deg(g):\ g\in \Ann(s)^\times\}.$$ 
\end{definition}
Let $\Min(s)$ denote the {\em set of MP's of $s$}. We do not require MP's to be monic.
The linear complexity (LC) of $s$ is $\LC(s)=\deg(f)\mbox{ where }f\in\Min(s).$
We will also write $\LC_n$ for $\LC(s)$ when $s$ is understood; similarly  $\LC_j=\LC(s^{(j)})$ for $1\leq j\leq n$. 
It is convenient to set $\LC_0=0$.

The following two functions simplify many statements in what follows.
\begin{definition} [Exponent Function] We define the exponent function $\ee:D[x]^\times \times(\{0\}\cup \N)\ra \Z$ $$\ee(f,n)=n+1-2\deg(f).$$
\end{definition}
When $f$ is understood, we write $\ee_n$ for $\ee(f,n)$; for example, we often write $\ee_n=n+1-2\LC_n$ if $f\in\Min(s)$.
We conventionally set $\ee_0=1$.

 \begin{definition}[Index Function]\label{indices}
Let  $s\in D^n$. We set $\mu^{(0)}=1$.
Given $\mu^{(j)}\in \Min(s^{(j)})$ and $\Delta_{j+1}=\Delta_{j+1}(\mu^{(j)})$ for $1\leq j\leq n-1$, put $0'=-1$ and define  $n'$ inductively by  
$$n'=\left \{\begin{array}{ll}
 \ n-1 & \mbox{ if } \Delta_{n}\neq 0\mbox{ and }\ee_{n-1}>0\\
(n-1)' &\mbox{ otherwise.}\\
\end{array}
\right.
$$
\end{definition}

 To simplify statements in the remainder of the paper, we will write $\mu'^{(j-1)}$ for $\mu^{((j-1)')}$ and $\Delta'_j$ for $\Delta(\mu'^{(j-1)},s^{((j-1)'+1)})$.

  The {\em polynomial part of $f$ and $s$}, written $[f\cdot s]$, is the $D[x]$-summand of $f\cdot\ul{s}$:
$$[f\cdot\ul{s}](x)=
\sum_{j=0}^{\vv(\ul{s})+d} (f\cdot \ul{s})_j\ x^j.$$

When $s$ is understood, we will write
$[f]$ for $[f\cdot\ul{s}]$, 

 \subsection{MP matrices}
\bd Let $s\in D^n$. We will call a $2\times 2$ matrix $M=M(s)$ with entries from $D[x]$ an {\em MP-matrix for $s$} if $M_{11}\in \Min(s)$, $M_{12}=[M_{11}]$, $M_{21}=M_{11}'$ and $M_{22}=[M_{21}]$.
\ed
When $s$ is understood, we will write $M^{(n)}$ for $M^{(n)}(s)$.
To express the main theorem in matrix terms, we need an integer Heaviside function
$$\theta(i)=\left\{\begin{array}{ll}
1 &\mbox{if } i>0\\
0 & \mbox{if } i\leq 0.\end{array}\right.
$$
Trivially, $e\cdot\theta(e)=\max\{e,0\}$. 
\bt   \label{bimrt}(Cf. \cite{Be68}, \cite{Ma69},  \cite{N95b}) 
Let  $s\in D^n$, $\varepsilon\in D$ be arbitrary but fixed. Put

$$M^{(0)}=\left[\begin{array}{lr}
1 & 0\\
\varepsilon & -1\end{array}\right]$$
and $\Delta_0=1$. 
Suppose that $M^{(j)}$ is an MP-matrix for $s^{(j)}$ for $1\leq j\leq n-1$ and $e=\ee_{n-1}$.

If $\Delta_n=0$ then $M^{(n-1)}$ is an MP-matrix for $s$, $\Delta'_{n+1}=\Delta'_n$ and $\ee_n=e+1$.

On the other hand, if $\Delta_{n}\neq 0$ then

(i) \ $\LC_n=\max\{e,0\}+\LC_{n-1}=n'+1-\LC_{n'}$

(ii)\ $U^{(n-1)}\cdot M^{(n-1)}$ is an MP-matrix for $s$
 where

$$U^{(n-1)}=\left[\begin{array}{cc}
\Delta'_{n}\cdot x^{e\cdot\theta(e)}& -\Delta_{n}\cdot x^{-e\cdot(1-\theta(e))}\\
\theta(e)	&1-\theta(e)\end{array} \right]$$

(iii)\ $|U^{(n-1)}|=\Delta'_{n+1}=\left\{\begin{array}{ll}
\Delta_n & \mbox{if } e>0\\
\Delta'_n & \mbox{if }e\leq 0\end{array} \right.$

(iv)\ $\ee_n=-|e|+1$

(v)\ if $M^{(n)}_{12}\neq 0$, $\deg(M^{(n)}_{12})=\max\{e,0\}+\deg(M^{(n-1)}_{12})$ .
\et
\be\label{rex} Let $r=(1,1,0,1,0,0)\in\F_2^6$ and $\varepsilon=0$. We have  $M^{(0)}=I$ by definition, $\ee_0=1$ and $\Delta_1=r_1=1$, so that $M^{(1)}=U^{(1)}M^{(0)}=\left[\begin{array}{ll}
x & 1\\
1 & 0
\end{array}
\right]=U$ say. Next, $\ee_1=0$ and $\Delta_2=r_2=1$, so that  $U^{(2)}= \left[\begin{array}{ll}
1 & 1\\
0 & 1\\ \end{array}
\right]
$  giving $$M^{(2)}=U^{(2)}M^{(1)}=\left[\begin{array}{ll}
1 & 1\\
0 & 1\\ \end{array}
\right]\left[\begin{array}{ll}
x & 1\\
1 & 0
\end{array}
\right]=
\left[\begin{array}{cc}
x+1 & 1\\
1 & 0
\end{array}
\right].$$ 
In the same way, $\ee_2=1=\Delta_3$ and
$$M^{(4)}=M^{(3)}=UM^{(2)}=\left[\begin{array}{cc}
x^2+x+1 & x\\
x+1 & 1 \end{array}
\right].
$$
\end{example}

We will also write
$M^{(n)}$ in terms of MP's as
$$M^{(n)}=\left[\begin{array}{l}
\ol{\mu}^{\ (n)}\\
\ol{\mu}'^{(n)}\end{array} \right]=\left[\begin{array}{ll}
\mu^{\ (n)} & [{\mu}^{\ (n)}]\\
\mu'^{(n)} & [{\mu}'^{(n)}]\end{array} \right]$$
where $\mu^{(n)}\in\Min(s)$ and $\ol{\mu}^{(n)}=(\mu^{(n)},[\mu^{(n)}])$.
In this formulation,  Theorem \ref{bimrt} yields $$\ol{\mu}^{(n)}=\left\{\begin{array}{ll}
\Delta'_{n}\cdot x^{\ e}\ \ol{\mu}^{(n-1)}-\Delta_{n}\cdot  \ol{\mu}'^{(n-1)}
&\mbox{ if }e> 0\\\\
\Delta'_{n}\cdot  \ol{\mu}^{(n-1)}-\Delta_{n}\cdot x^{-e}\ \ol{\mu}'^{(n-1)}& \mbox{ otherwise}
\end{array}\right.
$$
and we see again that there is a net increase of $e$ in LC precisely when $e>0$.
For the convenience of the reader we recall the algorithm implied by Theorem \ref{bimrt}; it is a rewrite of \cite[Algorithm 4.6]{N95b}.
 \begin{algorithm}\label{rewrite}  Algorithm MP ( Cf. \cite[p. 184]{Be68}), \cite[p. 124]{Ma69})

\begin{tabbing}
\noindent Input: \ \ \=$n\geq 1$, $\varepsilon\in D$ and $s=(s_1,\ldots,s_{n})\in D^n$.\\

\noindent Output: \>$\ol{\mu}=(\mu,[\mu])$.\\

\{$e := 1$;\ $\ol{\mu}':=(\varepsilon,-1)$;\ $\Delta':=1$;
$\ol{\mu}  :=  (1,0);\ $\\
{\tt FOR} \= $j = 1$ {\tt TO }$n$\\
    \> \{$\Delta    :=  \sum_{k=0}^{\frac{j-e}{2}} \mu_k \  s_{k+\frac{j+e}{2}};$ \\
   \> {\tt IF} \=$\Delta  \neq  0$\\ 
   \>\>{\tt THEN}\{{\tt IF} $e\leq 0$  \=   						{\tt THEN} $\ol{\mu}   :=  \Delta'\cdot \ol{\mu}-\Delta\cdot  x^{-e} \ol{\mu}'$;\\\\
  \>                 \>   \> {\tt ELSE} \{\=$(t,u) := \ol{ \mu}$;\\
  \>                 \>    \>       \> $\ol{\mu}  :=   \Delta'\cdot x^e\ol{\mu}-\Delta\cdot  \ol{\mu}'$;\\
  \>                 \>     \>       \> $\ol{\mu}':= (t,u)$; \ $\Delta':= \Delta$;\\ 
    \>                 \>    \>        \>$e := -e$\}\}\\
  \> $e  := e+1$\}\\
{\tt RETURN}$(\ol{\mu})$\}\\
\end{tabbing} 
\end{algorithm}
The analogue of $M^{(n)}$ in Berlekamp's context and notation  is
$\left[\begin{array}{ll}
\sigma &\omega\\
\tau & 
\gamma\end{array}\right]$, \cite[p. 181]{Be68}.
See also \cite[p. 180]{Bl83}. We derive our analogue as follows. Let $\pp:\{0,\ldots,n-1\}\rightarrow\{0,\ldots,n-1\}$
be defined by $$\pp(j)=\left\{\begin{array}{ll}
1 &\mbox{ if } j=0\\
j-j' & \mbox{ otherwise.}\end{array}\right.
$$
Then $\pp(n)=\pp(n-1)+1$  if $\Delta_n=0$. Let  $\mu^{(n)\ast}$ and $\mu'^{(n)\ast}$ denote the reciprocal polynomials. Considering the cases $e>0$ and $e\leq 0$ gives the following corollary of Theorem \ref{bimrt}.
\begin{corollary} \label{recip}
If $\Delta_{n}\neq 0$ then

(i) $\mu^{(n)\ \ast}=\Delta'_{n}\cdot \mu^{(n-1)\ \ast}- \Delta_{n} \cdot
x^{\pp(n-1)}\ \mu'^{(n-1)\ \ast} $
  
(ii) $\pp(n)=\pp(n-1)+1$ if $\ee_{n-1}\leq 0$ and $\pp(n)=1$  otherwise.
\end{corollary}
 
In this way, we obtain a linear feedback shift-register  of shortest length $\LC=\frac{n+1-e}{2}$  and 'feedback polynomial' $\mu^{(n)\ast}$ generating $s^{(n)}$. 
Our corresponding updating matrix is
$$\left[\begin{array}{ll}
\Delta'_{n}& -\Delta_{n}\cdot x^{p}\\
\theta(e)	&1-\theta(e)\end{array} \right]$$
where $p=\pp(n-1)$, $\pp(0)=0$,  $\pp(n)$ is set to 0 if $e>0$  and then $\pp(n)=p+1$ (regardless of $e$). 
\subsection{B\'ezout Identities}
\bd For $s\in D^n$, we set $\nabla_0=1$ and 
$$\nabla_n=\left \{
\begin{array}{ll}
	\nabla_{n-1} &\mbox{ if }\Delta_{n}=0\\\\
	\Delta'_{n+1}\nabla_{n-1}&\mbox{ if } \Delta_{n}\neq 0.\\
\end{array}
\right.
$$
\ed
 \bp\label{numu}(Cf. \cite[Theorem 7.42]{Be68}) $|M^{(0)}|=-\nabla_0$ and
 $|M^{(n)}| =\nabla_n.$
\ep
\begin{proof} The first statement is trivial. Inductively, if $\Delta_n=0$, there is nothing to prove; otherwise Theorem \ref{bimrt} gives  $$|M^{(n)}|=|U^{(n-1)}|\cdot|M^{(n-1)}|=\Delta'_{n+1}\cdot|M^{(n-1)}|.$$
\epr
Thus we have $n$ B\'ezout identities for $s$: for $1\leq j\leq n$
\begin{equation}\label{bezout}\mu^{(j)}\ [\mu'^{(j)}] - [\mu^{(j)}]\ \mu'^{(j)}
 =\nabla_j
 \end{equation} proved from first principles in \cite[Theorem 3.3]{N2010c}.  Again we see that $\gcd(\mu^{(j)},[\mu^{(j)}])=1$ if $D$ has unique factorisation; likewise, $\gcd(\mu^{(j)},\mu'^{(j)})=1$.
 
 \begin{table}\label{rueppeltable}
\caption{Algorithm MP $\varepsilon=0$,   
$r=(1,1,0, 1,0,0)\in \F_2^{6}$.}
\begin{center}
\begin{tabular}{|c|r|r|l|l|l|l|}\hline
$j$ & $\Delta_j$  & $\ee_{j-1}$ 	&$\mu^{(j)}$ &$\mu'^{(j)}$\\\hline\hline
$0$   &$1$   &$ $  & $1$ & $0$  \\\hline
$1$   &$1$  &$0$  & $x$ &$1$ \\\hline
$2$   & $1$ &$1$ & $x+1$ & $1$  \\\hline
$3$   & $1$ &$0$  & $x^2+x+1$ & $x+1$  \\\hline
$4$   &$0$  &$1$  & $x^2+x+1$ & $x+1$  \\\hline
$5$   & $1$ &$0$  & $x^3+x^2+1$  & $x^2+x+1$ \\\hline
$6$   &$0$  &$1$  & $x^3+x^2+1$ &  $x^2+x+1$ \\\hline
\end{tabular}
\end{center}
\end{table}

 \section{PLCP Sequences}
The following is a slight generalisation of \cite{Rueppel}:  $s\in D^n$ has a {\em perfect linear-complexity profile (PLCP)} if   $\LC_j = \lfloor \frac{j+1}{2}\rfloor$ for $1\leq j\leq n$. If $s:\N\ra D$, then $s$ has a PLCP if for all $n$, $s^{(n)}$ has a PLCP.
It is easy to see that the binary  sequences of length 1 to 4  with a PLCP are $(1)$, $(1,s_2)$, $(1,1,0), (1,0,1)$ and $(1,1,0,s_4), (1,0,1,s_4)$; see Table \ref{plcp} for their $\ol{\mu}^{(j)}$.

 \subsection{Basic Characterisations}\label{basics}
 Recall that for any sequence, $\mu^{(0)}=1$ and $\ee_0=1$.

\bp \label{basictfae}\tfae

(i) $s$ has a PLCP

(ii) $\LC_1=1$ and  for $2\leq j\leq n$
$$\LC_j-\LC_{j-1}= \left\{\begin{array}{rl}

         0& \mbox{if } j \mbox{ is even}\\
	1 &\mbox{otherwise,}
 \end{array}
\right. $$

(iii) for $1\leq j\leq n$
$$\ee_j= \left\{\begin{array}{rl}

         1& \mbox{if } j \mbox{ is even}\\
	0 &\mbox{otherwise,}
 \end{array}
\right. $$

(iv) $\Delta_{j}\neq 0$ for all odd $j$, $1\leq j\leq n$\\

(v)  for $2\leq j\leq n$, 
$$(j-1)'=\left\{ \begin{array}{ll}
j-2 &\mbox{ if } j-1\mbox{ is even}\\
j-3 & \mbox{ otherwise,}
\end{array}
\right.
$$

(vi) $\ol{\mu}^{(1)}=(x+\varepsilon,1)$ and for $2\leq j\leq n$
$$\ol{\mu}^{(j)}= \left\{\begin{array}{ll}
        \Delta_{j-1}\cdot \ol{\mu}^{(j-1)}-\Delta_j\cdot\ol{\mu}^{(j-2)}&\mbox{ if } j \mbox{ is even}\\\\
\Delta_{j-2}\cdot x\ol{\mu}^{(j-1)}-\Delta_{j}\cdot\ol{\mu}^{(j-3)}& \mbox{ otherwise.}
 \end{array}
\right. $$
\ep
\begin{proof}
(i) $\Leftrightarrow$ (ii) $\Leftrightarrow$ (iii): Easy consequence of the definitions.

(i) $\Rightarrow$ (iv): If $j\leq n+1$ is odd then $\Delta_j\neq 0$, for otherwise $\frac{j-1}{2}+1=\frac{j+1}{2}=\LC_{j}=\LC_{j-1}=\frac{j-1}{2}$. 

(iv) $\Rightarrow$ (i): Let  $\Delta_{j}\neq 0$ for all odd $j$, $1\leq j\leq n+1$. Then $s_1\neq 0$, $\LC_1=1$ and $\ee_1=0$. If $\Delta_2=0$, then $\LC_2=\LC_1=1$, otherwise $\LC_2=\max\{\ee_1,0\}+1=1$, so that $\LC_2$ is as required. Suppose that $j\leq n$ is odd and $\LC_{k}=\lfloor\frac{k+1}{2}\rfloor$ for all $k$, $1\leq k\leq j-1$. We have $\LC_j=j-\LC_{j-1}=j-\frac{j-1}{2}=\lfloor\frac{j+1}{2}\rfloor$. If $j=n+1$, we are done. Otherwise,  if $\Delta_{j+1}=0$, we have $\LC_{j+1}=\LC_j=\lfloor\frac{j+1}{2}\rfloor=\lfloor\frac{j+2}{2}\rfloor$, whereas if $\Delta_{j+1}\neq 0$,
$\LC_{j+1}=j+1-\LC_j=j+1-\lfloor\frac{j+1}{2}\rfloor=\lfloor\frac{j+2}{2}\rfloor$.

(iii) $\Ra$ (v): Let $i=(j-1)'$. We have 

If $j$ is even then $\Delta_{j-1}\neq 0$ and $\ee_{j-2}=1$, so $i=j-2$; if $j$ is odd then $\ee_{j-2}=0$ and $\ee_{j-3}=1$ so $i=(j-2)'=j-3$. 
(Note that $i+1$ is always odd, so that $\Delta_{i+1}\neq 0$.)

(v) $\Ra$ (vi): Inserting $(j-1)'$ and $\ee_{j-1}$ in Theorem \ref{bimrt} gives the formulae for $\ol{\mu}$.

(vi) $\Ra$ (iii): We have $\LC_1=1$, $\LC_j=\LC_{j-1}$ if $j$ is even and $\LC_j=\LC_{j-1}+1$ if $i$ is odd. Thus if $j$ is odd,
$\ee_j=j+1-2\LC_j=j+1-2(\LC_{j-1}+1)=j+1-2\LC_{j-1}=\ee_{j-1}+1$. Applying this inductively gives (iii).
\epr

 \begin{table}   \label{plcp}

\caption{\vspace{0.25cm}$\ol{\mu}^{(j)}\in \F_2[x]^2$ for $0\leq i\leq 4$.}
\begin{center}
\begin{tabular}{|r|l|}\hline
$j$ & $\ol{\mu}^{(j)}$\\\hline\hline
$0$ & $(1,0)$\\\hline
$1$ & $(x,1)$\\\hline
$2$ & $(x+\Delta_1,1)$\\\hline
$3$ & $(x^2+\Delta_1 x+1,x)$\\\hline
$4$ & $(x^2+(\Delta_1+\Delta_3)x+1,x+\Delta_3)$\\\hline
\end{tabular}
\end{center}
\end{table}

Thus without loss of generality, we may assume that $n$ is odd, and it is easy to see that for odd $n$, ${\mu}^{(n)}+c\cdot {\mu}'^{(n)}={\mu}^{(n)}+c\cdot {\mu}^{(n-1)}\in \Min(s)$ for any $c\in D$.

\bc Let $D=\F$ be a field, $s\in D^n$, ${\mu}^{(j)}$  and $\Delta_j$ be as in Theorem \ref{bimrt} for $1\leq j\leq n$. If $s$ has a PLCP and $\mu^{(j)}\in\Min(s)$, then
$$\mu^{(j)}= \left\{\begin{array}{ll}
{\mu}^{(j-1)}-\frac{\Delta_j}{\Delta_{j-1}}\cdot{\mu}^{(j-2)} 
            &  \mbox{ if } j \mbox{ is even}\\\\
(x+c)\cdot{\mu}^{(j-1)}-\frac{\Delta_j}{\Delta_{j-2}}\cdot{\mu}^{(j-3)} &\mbox{ otherwise}
 \end{array}
\right. $$
where $c\in D$.
\ec

\begin{proof} If $j$ is even, then $s^{(j)}$ has a unique monic MP and if $j$ is odd,  any monic MP of $s^{(j)}$ is 
$${\mu}^{(j)}+c\cdot{\mu}'^{(j)}
=(x{\mu}^{(j-1)}-\frac{\Delta_j}{\Delta_{j-2}}\cdot{\mu}^{(j-3)})+c\cdot{\mu}^{(j-1)}
=(x+c)\cdot{\mu}^{(j-1)}-\frac{\Delta_j}{\Delta_{j-2}}\cdot{\mu}^{(j-3)}$$ for some $c\in D$ by \cite[Theorem 4.16]{N95b}.
\epr

\subsection{The Characterisation of Wang and Massey}
\label{WM}
Here we prove a theorem of  Wang and Massey \cite{Wangpdf} on binary sequences using the B\'ezout identities (\ref{bezout}) and results of Section \ref{basics}. Thus $D=\F_2$ throughout this subsection.
Let us call $s\in D^n$ {\em stable} if $s_1=1$ and for even $j$, $2\leq j\leq n$, $s_{j+1}=s_j+s_{\frac{j}{2}}$.

The transform  $\ul{t}$ which appears in the next result was used in  \cite[Theorem 3]{Nied87}  and a similar one was used in \cite{Wangpdf}.
\bp \label{ult} 
Let $n$ be odd and $\ul{t}=\ul{s}^2+(x+1)\ul{s}+1$. Then $s$ is stable if and only if $t_j=0$ for $j$ even, $0\leq j\leq n$.
\ep
\begin{proof} We have $t_0=s_1+1$ and  ${t}_j=s_j+s_{j+1}+s_{\frac{j}{2}}$ for all even $j$, $2\leq j\leq {n-1}$.\\
\epr

 \begin{table}   \label{plcp}
\caption{\vspace{0.25cm}${\sigma}^{(j)}\in \F_2[x]$ for $0\leq i\leq 4$.}
\begin{center}
\begin{tabular}{|r|l|}\hline
$j$ & $\sigma^{(j)}$\\\hline\hline
$0$ &$1$\\\hline
$1$ &$x+1$\\\hline
$2$ &$(\Delta_1+1)x+1$\\\hline
$3$ &$(\Delta_1+1)x^3+x+1$\\\hline
$4$ &$(\Delta_1+1)x^3+(\Delta_1\Delta_3+1)x+1$\\\hline
\end{tabular}
\end{center}
\end{table}
 \bl (Cf. \cite[Lemma 1]{Wangpdf})\label{sigma} Let
 $\sigma^{(j)}=[\mu^{(j)2}\cdot \ul{t}]$. If $s$ has a PLCP then for $2\leq j\leq n$,
 $$\sigma^{(j)}=\left\{\begin{array}{ll}
\sigma^{(j-1)}+\Delta_j\cdot \sigma^{(j-2)}
+\Delta_j\cdot (x+1)& \mbox{if } j 	      \mbox{ is even}\\\\
x^2\sigma^{(j-1)}+\sigma^{(j-3)}+x(x+1) &\mbox{otherwise.}
\end{array}
\right.
$$ 
\el
\begin{proof} From the definition, $$\sigma^{(j)}=(x+1)\mu^{(j)}[\mu^{(j)}]+\mu^{(j)2}+[\mu^{(j)}]^2.$$ 
Let $j$ be even, so that  $\ol{\mu}^{(j)}=\ol{\mu}^{(j-1)}+\Delta\cdot\ol{\mu}^{(j-2)}$ by Theorem \ref{basictfae}, where $\Delta=\Delta_j$. Putting $\nu^{(i)}=[\mu^{(i)}]$ and expanding $\sigma^{(j)}$ gives

$$\sigma^{(j-1)}+\Delta\sigma^{(j-2)}+\Delta(x+1)(\nu^{(j-2)}\mu^{(j-1)}+\mu^{(j-2)}\nu^{(j-1)})$$
which is as required since $\nu^{(j-2)}\mu^{(j-1)}+\mu^{(j-2)}\nu^{(j-1)}=1$ by Proposition \ref{numu}. The proof for $j$ odd is similar.
\epr
The reader may also check that $\deg(\sigma^{(j)})=
 j-1$ if $j$ is even and $j$ if $\deg(\sigma^{(j)})=j$ is odd, but we will not need this.

\bt \label{link} Let  $n\geq 1$ be odd and $s\in D^n$. If $s$ has a PLCP then $s$ is stable. 

\et
\begin{proof} An easy inductive proof using Lemma \ref{sigma} shows that if $s$ has a PLCP, then for $0\leq j\leq n$, $\sigma^{(j)}_0=1$ and $\sigma^{(j)}_2=0$
  for $2\leq j\leq n$. Let $j$ be even with $0\leq j\leq n-1$. As $t_0=0$, we can assume inductively that $t_j=0$ for even $j$, $2\leq j\leq n-3$.  Put $\mu=\mu^{(n)}$, which has degree $\frac{n+1}{2}$. Now
$$\sigma^{(n)}=[\mu^2\cdot\ul{t}]=[\mu^2\cdot(t_3x^{-3}+\cdots +t_{n-1}x^{1-n}+t_nx^{-n})].$$
As $\sigma^{(n)}_2=0$, the quadratic term  in the right-hand side viz. $\mu_{\frac{n+1}{2}}\cdot t_{n-1}$ is zero. This  forces $t_{n-1}=0$ as $\mu_{\frac{n+1}{2}}$ is the leading coefficient of $\mu$, and $s$ is stable by Proposition \ref{ult}.
\epr
\bp\label{count}  The number of $s\in \F_q^n$ which have a PLCP is $(q-1)^{\lceil \frac{n}{2}\rceil}q^{\lfloor \frac{n}{2}\rfloor}$. \ep
\begin{proof} A sequence $s$ determines $(\Delta_1,\ldots,\Delta_{n})\in D^n$ uniquely, and conversely. Thus the result follows from Proposition \ref{basictfae}(iv).
\epr
\bc  \label{stablewm}(\cite[Theorem p.13]{Wangpdf})
Let  $n\geq 1$ be odd and $s\in D^n$. Then $s$ has PLCP if and only $s$ is stable.
\ec  \begin{proof} As in noted in \cite{Wangpdf}, there are clearly $\lceil \frac{n}{2}\rceil$ stable sequences in $D^n$, so the result follows from Proposition \ref{count}.
\epr

For other proofs of the characterisation of Wang and Massey, see \cite[Corollary 1]{Nied86a} (which uses the characterisation of \cite{BaumSweet}) and \cite[Theorem 3]{Nied87} (which   uses an idea of  \cite{Taussat}).
\subsection{Rueppel's Conjecture}

Throughout this section, $r:\N\ra \F_2$ denotes the binary sequence with $r_{2^k}=1$  for $k\geq 0$ and zero otherwise.  We have already seen the two invertible matrices
$$U=\left[\begin{array}{ll}
x & 1\\
1 &	0\end{array} \right]\mbox{ and }M=M^{(2)}(r)=
\left[\begin{array}{cl}
x+1 & 1\\
1 & 0
\end{array}
\right]$$
as well as  powers of $U$ in Example \ref{rex}.  Our goal in this section is to show that the updating matrix of $r$ is $U$ if $n$ is even and the $2\times 2$  identity matrix if $n$ is odd. 

To work with the powers of $U$, we define a sequence of binary polynomials as follows: $\gamma^{(0)}=0$, $\gamma^{(1)}=1$ and 
$$\gamma^{(k)}=x\gamma^{(k-1)}+\gamma^{(k-2)}\mbox{ for }k\geq 2.$$

The first eight are $0,1,x,x^2+1,x^3,x^4+x^2+1,x^5+x$ and $ x^6+x^4+1$.  
A simple induction gives the powers of $U$ in terms of the polynomials $\gamma^{(i)}$.
\bp \label{Upowers}For $k\geq 1$, $$U^k=
		\left[\begin{array}{cc}
\gamma^{(k+1)} & \gamma^{(k)}\\
\gamma^{(k)}  & \gamma^{(k-1)}\\
\end{array}
\right].$$
\ep

\bl\label{gamma}(Cf. \cite[Lemma 1]{Dai})

(i) For $m\geq n$, $\gamma^{(m+n)}=x\gamma^{(m)}\gamma^{(n)}+\gamma^{(m-n)}$. In particular, $\gamma^{(2n)}=x\gamma^{(n)2}$.

(ii) $\deg(\gamma^{(n)})=n-1$.

(iii) $(\gamma^{(n)}+\gamma^{(n-1)})(0)=1$.

(iv)  $\gcd(\gamma^{(n)},\gamma^{(n-1)})=1$.
\el
\begin{proof} (i) The case $n=1$ is the definition. Suppose inductively that the result is true for $n-1\leq m$. We proceed as follows:
\begin{eqnarray*}
\gamma^{(m+n)}&=&x\gamma^{(m+n-1)}+\gamma^{(m-n-2)}
=x\left[x\gamma^{(m)}\gamma^{(n-1)}+\gamma^{(m-n+1)}\right]+\gamma^{(m+n-2)}\\
&=&x\gamma^{(m)}\left[x\gamma^{(n-1)}+\gamma^{(n-2)}\right]+x\gamma^{(m)}\gamma^{(n-2)}+x\gamma^{(m-n+1)}+\gamma^{(m+n-2)}\\
&=&x\gamma^{(m)}\gamma^{(n)}+x\gamma^{(m)}\gamma^{(n-2)}
+x\gamma^{(m-n+1)}+\gamma^{(m+n-2)}
=x\gamma^{(m)}\gamma^{(n)}+\gamma^{(m-n)}\end{eqnarray*}
since if $m=n$, $x\gamma^{(n)}\gamma^{(n-2)}+x\gamma^{(1)}+\gamma^{(2n-2)}=\gamma^{(2n-2)}+\gamma^{(2)}+x\gamma^{(1)}+\gamma^{(2n-2)}=0=\gamma^{(m-n)}
$
and we are done. Otherwise the inductive hypothesis yields
\begin{eqnarray*}
x\gamma^{(m)}\gamma^{(n-2)}=\gamma^{(m+n-2)}+\gamma^{(m-n+2)}=\gamma^{(m+n-2)}+x\gamma^{(m-n+1)}+\gamma^{(m-n)}.
\end{eqnarray*}

The remaining items are easy inductions.
\epr

\bp \label{sum} For $k\geq 0$, (i)  $\gamma^{(2^k)}=x^{2^k-1}$ and (ii) $\gamma^{(2^k-1)}=\sum_{j=1}^kx^{2^k-2^j}$.
\ep
\begin{proof} (i) The result is true for $k=0$. The rest of the proof is a simple induction using $m=n=2^{k-1}$ in Lemma \ref{gamma}(i). The proof of Part (ii) is similar to \cite[Lemma 1(3)]{Dai}. The result is true for $k=0$, so suppose inductively that it is true for $k-1$. Since $2^k-2^j=2^j+\cdots +2^{k-1}$, we have\begin{eqnarray*}
\sum_{j=1}^kx^{2^k-2^j}
=1+\sum_{j=1}^{k-1}x^{2^k-2^j}=1+\sum_{j=1}^{k-1}\prod _{i=j}^{k-1}x^{2^i}
=1+\sum_{j=1}^{k-1}\prod _{i=j}^{k-1}x\gamma^{(2^i)}
\end{eqnarray*}
We claim that the right-hand side is $\gamma^{(2^{k}-1)}$. This is true for $k=1$ and assuming that it is true for $k-1$, the right-hand side is
\begin{eqnarray*}1+\left[1+\sum_{j=1}^{k-1}\prod _{i=j}^{k-2}x\gamma^{(2^i)}\right]\cdot x\gamma^{(2^{k-1})}=1+\gamma^{(2^{k-1}-1)}\cdot x\gamma^{(2^{k-1})}=1+(\gamma^{(2^{k}-1)}+1)
\end{eqnarray*}
where the last equality follows from Lemma \ref{gamma}(i).
\epr

We have
$$\ul{r}^{(n)}=\sum_{i=0}^{\lfloor \log_2 n\rfloor} x^{-2^i}.$$
The next lemma is essentially \cite[Lemma 2]{Dai} with a simpler proof. It is key to determining the discrepancies.
\bl\label{r0} For $k\geq 1$
$$\left[\begin{array}{c}
\ul{r}^{(2^k)}\\
1\end{array}
\right]=x^{-2^k}M^{-1}U^{2-2^k}\left[\begin{array}{c}
1\\
x+1
\end{array}
\right].
$$
\el
\begin{proof} For $p\geq 2$ direct evaluation yields
$$M^{-1}U^{2-p}\left[\begin{array}{c}
1\\
x+1
\end{array}
\right]=\left[\begin{array}{c}
\gamma^{(p-1)}+\gamma^{(p)}\\
\gamma^{(p+1)}+\gamma^{(p-1)}\end{array}
\right].
$$
Proposition \ref{sum} with $p=2^k$ implies that $$\gamma^{(2^k-1)}+\gamma^{(2^k)}=x^{2^k-1}+\sum_{j=1}^kx^{2^k-2^j}=x^{2^k}\ul{r}$$ and $\gamma^{(2^k+1)}+\gamma^{(2^k-1)}=x\gamma^{(1)}\gamma^{(2^k)}=x^{2^k}.$ 
\epr
Now we can state and prove the main result of this section.\\

\bt\label{rueppelconj} (Cf. \cite{Dai}) If $M^{(n)}=M^{(n)}(r)$ then $M^{(2)}=M$ and for $n\geq 3$, $$M^{(n)}=\left\{\begin{array}{ll}
M^{(n-1)}&\mbox{ if }n \mbox{ is even}\\
UM^{(n-1)}=U^{\frac{n-1}{2}}M&\mbox{ if }n \mbox{ is odd.}
\end{array}
\right.
$$
\et
\begin{proof} The matrix $M=M^{(2)}$ and  the result for $n=3,4$ were derived in Example \ref{rex}. Suppose inductively that $n$ is odd and the result is true for sequences of length $n-1\geq 4$. First we use Lemma \ref{r0} to show that  $\mu^{(n-1)}=\mu^{(n-2)}\not\in\Ann(r^{(n)})$. Let $k=\lfloor \log_2 n\rfloor$ so that $\ul{r}^{(n)}=\ul{r}^{(2^k)}$.
Lemma \ref{r0} and the inductive hypothesis give\begin{eqnarray*}
M^{(n-2)}\left[\begin{array}{c}
 \ul{r}^{(n)}\\
1\end{array}
\right]
=U^{\frac{n-3}{2}}M\left[\begin{array}{c}
 \ul{r}^{(2^k)}\\
1\end{array}
\right]
=U^{-p}x^{-2^k}\left[\begin{array}{c}
1\\
x+1
\end{array}
\right]
=x^{-2^k}
		\left[\begin{array}{cc}
\gamma^{(p-1)} & \gamma^{(p)}\\
\gamma^{(p)}  & \gamma^{(p+1)}\\
\end{array}
\right]\left[\begin{array}{c}
1\\
x+1
\end{array}
\right]
\end{eqnarray*}
 where $p=2^k-\frac{n+1}{2}\geq 0$. As $\LC_{n-2}=\frac{n-1}{2}$, the discrepancy 
$(\mu^{(n-2)}\cdot \ul{r}^{(n)})_{\frac{n-1}{2}-n}$ is
$$\left((x+1)\gamma^{(p)}+\gamma^{(p-1)}\right)_{p}+\gamma^{(\frac{n-1}{2})}_{\ \ \ \frac{n-1}{2}-n}$$
which is 1 since  $\deg(\gamma^{(p)})=p-1$, $\deg(\gamma^{(p-1)})=p-2$ and $n>1$. 
Next we construct $M^{(n)}$. As $\LC_{n-1}=\LC_{n-2}=(n-1)/2$, $\ee_{n-1}=n-2\LC_{n-1}=1$ and $M^{(n)}=UM^{(n-2)}=U^{(\frac{n-1}{2})}M$
 by Theorem \ref{bimrt} and the inductive hypothesis.  

It remains to show that ${\mu}^{(n)}\in\Ann(r^{(n+1)})$. If $n+1<2^{k+1}$ then  $\ul{r}^{(n+1)}=\ul{r}^{(2^k)}$ and applying the first part gives 
\begin{eqnarray*}
M^{(n)}\left[\begin{array}{c}
 \ul{r}^{(n)}\\
1\end{array}
\right]
=UM^{(n-2)}
\left[\begin{array}{c}
 \ul{r}^{(2^k)}\\
1\end{array}
\right]=x^{-2^k}\left[\begin{array}{cc}
\gamma^{(p-2)} & \gamma^{(p-1)}\\
\gamma^{(p-1)}  & \gamma^{(p)}\\
\end{array}
\right]
\left[\begin{array}{c}
1\\
x+1
\end{array}
\right].
\end{eqnarray*} So the discrepancy 
$(\mu^{(n)}\cdot \ul{r}^{(n+1)})_{\frac{n+1}{2}-n-1}$ is
$\left((x+1)\gamma^{(p-1)}+\gamma^{(p-2)}\right)_{p+1}+\gamma^{(\frac{n+1}{2})}_{\ \ \frac{n+1}{2}-n-1}=0.$
If $n+1=2^{k+1}$ then
\begin{eqnarray*}
M^{(n)}\left[\begin{array}{c}
 \ul{r}^{(n+1)}\\
1\end{array}
\right]
=U^{-p}x^{-n-1}\left[\begin{array}{c}
1\\
x+1
\end{array}
\right]=\left[\begin{array}{cc}
\gamma^{(p-1)} & \gamma^{(p)}\\
\gamma^{(p)}  & \gamma^{(p+1)}\\
\end{array}
\right]x^{-n-1}\left[\begin{array}{c}
1\\
x+1
\end{array}
\right]
\end{eqnarray*}
where $p=\frac{n-1}{2}$. Then  
$\left(\mu^{(n)}\cdot \ul{r}^{(n+1)}\right)_{\frac{n+1}{2}-n-1}=\left((x+1)\gamma^{(p)}+\gamma^{(p-1)}\right)_{\frac{n+1}{2}}+\gamma^{(\frac{n+1}{2})}_{\ \ \frac{n+1}{2}-n-1}$
which is zero as before. Hence  $\mu^{(n)}\in\Min(r^{(n+1)})$ and the proof is complete.
\epr
\bc\label{mu}
For the sequence $r$, $\ol{\mu}^{(n)}$ is
$$\left\{\begin{array}{ll}
\ol{\mu}^{(n-1)}&\mbox{ if } n \mbox{ is even}\\
x\ol{\mu}^{(n-2)}+\ol{\mu}^{(n-4)}=(\gamma^{(p)}+\gamma^{(p-1)},\gamma^{(p-1)})&\mbox{ if } n \mbox{ is odd}
\end{array}\right.
$$
where $p=\frac{n+3}{2}$ if $n$ is odd and $\LC_n=\lfloor\frac{n+1}{2}\rfloor$.
\ec
\bc (Cf. \cite{Dai}) If $\mu^{(n)}$ is as in Corollary \ref{mu}, its reciprocal defines a  linear feedback shift-register of shortest length $\lfloor \frac{n+1}{2}\rfloor$ which generates $r^{(n)}$.
\ec

\brs (i) Let $Y^2+xY+1\in \F_2(x)[Y]$ have roots $\rho,\rho^{-1}$ in some algebraic extension of $\F_2(x)$ and $\theta_n=\rho^n+\rho^{-n}$, as in \cite{Dai}. Parts 1,2 of \cite[Lemma 1]{Dai} easily imply that  for $n\geq 1$, $x\gamma^{(n)}(x)=\theta_n(x)$. Lemma \ref{r0}  trivially implies \cite[Lemma 2]{Dai} and hence \cite[Lemma 3]{Dai}. Further, \cite[Lemma 4]{Dai} is a trivial consequence of Proposition \ref{Upowers} and \cite[Lemma 5]{Dai} is immediate. It follows that the main results of \cite{Dai} can be proved using the binary polynomials $\gamma^{(n)}$ and the Euclidean algorithm, i.e. without introducing $Y^2+xY+1\in \F_2(x)[Y]$ and its roots.

(ii) It is not hard to show that if $s:\N\ra \F$ is a linear recurring sequence and $f\in \Ann(s)$, then $f\in\Min(s)$ if and only if $\gcd(f,[f])=1$. Also, if $s\in \F^n$ and $f\in\Min(s)$, then $\gcd(f,[f])=1$; apply Proposition \ref{numu} or see \cite[Corollary 3.24]{N95b}.  We note here that $r^{(n)}$ shows that the converse fails for finite sequences. Let $2^k\leq n<2^{k+1}$. Then  $$x^{2^k}\ul{r}^{(n)}=x^{2^k}\ul{r}^{(2^k)}=[x^{2k}]$$
so that $x^{2^k}\in\Ann(r^{(n)})$ and  $\gcd(x^{2^k},[x^{2^k}])=1$, but if $n$ is even or $n+1<2^{k+1}$ then $\LC_n(r)=\lfloor\frac{n+1}{2}\rfloor<2^k$ and $x^{2^k}$ is not an MP of ${r}^{(n)}$; cf. \cite[p. 230]{Nied86a}. It would be interesting to know when $f\in \Ann(s)$ and $\gcd(f,[f])=1$ implies $f\in\Min(s)$.

(iii) As noted in \cite[p. 231]{Nied86a}, the theorem of Wang and Massey also shows that $r$ has a PLCP.
\ers
\subsection{The Height of a Sequence}\label{hgt}
Let $D=\F$ be a field, $s:\N\ra \F$ and let $\{A_n\in\F[x]\}$ be the partial quotients in the continued fraction expansion of $\ul{s}$. Then $K(\ul{s})$ was defined in \cite[p. 223]{Nied86a} by 
$$K(\ul{s})=\sup_{n\geq 1}\{\deg(A_n)\}.$$ 

The next theorem  was proved  using inequalities satisfied by $K(s)$   in \cite[Theorem 2]{Nied86b} for $\F_q$-sequences. 
(As in \cite{Nied86a} {\em et seq.}, $s:\N\ra\F$ is {\em irrational} if it is not a linear recurring sequence.)
\bt \label{CFK} If $s:\N\ra \F$ is irrational then $$K(\ul{s})=\sup\{\ee_n\}.$$
\et
As $\{\ee_n\}$ figures prominently in Theorem \ref{bimrt}, we will take their maximum as our starting point. In general, the range of $\LC(s)$ gives $1-n\leq \ee_n\leq n+1$.
\bd[Height]  If $s\in D^n$, we set   
$$\hgt(s)=\max\{\ee_j:\ 1\leq j\leq n\}$$
and if $s:\N$ $\ra D$, we put $\hgt(s)=\lim_{n\ra \infty}\hgt(s^{(n)})$.
\ed

As $\hgt(s^{(n)})\leq \hgt(s^{(n+1)})$, the limit always exists, although it may be infinite. For example, if $s=(0,\ldots,0)\in D^n$, then $\hgt(s)=n+1$, so that if $s$ is the infinite zero sequence, $\hgt(s)=\infty$. We have  $\ee_1=2$ if $s_1=0$ and $\ee_1=1$ otherwise, so that $\hgt(s)\geq 1$.

 The terminology 'height' was suggested by Theorem(i) \ref{bimrt}: for $s\in D^n$, $\LC_n$ increases (by $e$) exactly when $e>0$. Thus $\hgt(s)$ is the maximum of the degree jumps in $s$, and we can compute it  using Algorithm MP or the Berlekamp-Massey algorithm. Since LC is non-decreasing, $\lim_{n\ra\infty}\LC_n$ exists.
 
 Theorem \ref{rueppelconj} immediately implies the next result.
 \bc(Cf. \cite{Nied86a}) $\hgt(r^{(n)})=1$.
\ec
\bp\label{bound}(Cf. \cite[Theorem 3]{Nied86a}) 

(i) If $s\in D^n$ then $\hgt(s^{(j)})\geq \ee_j\geq 1-\hgt(s^{(j)})\mbox{ for }1\leq j\leq n$.

(ii) If $s:\N$ $\ra D$ then $\hgt(s)\geq \ee_n\geq1- \hgt(s)$.
\ep
\begin{proof} The inequality $\ee_j\leq \hgt(s^{(j)})$ is trivial. We  prove that $1-\hgt(s^{(j)})\leq \ee_j$ by induction, the case $n=1$ being a trivial verification. Suppose that $n\geq 2$ and the result is true for all sequences of length $n-1$. If $\Delta_n=0$ or ($\Delta_n\neq 0$ and $\ee_{n-1}\leq 0$), then $\ee_n=\ee_{n-1}+1\geq (1-\hgt(s^{(n-1)}))+1\geq 1-\hgt(s)$ since $\hgt(s^{(n-1)})\leq \hgt(s)$. If 
$\Delta_n\neq 0$, then $\ee_n=-\ee_{n-1}+1\geq -\hgt(s^{(n-1)})+1\geq 1-\hgt(s)$.
Now let $s:\N$ $\ra D$. If $\hgt(s)=\infty$, the result is trivially true, otherwise taking limits gives the  required result.
\epr
If $s$ has a PLCP then $\hgt(s)=1=\ee_{2n}$ and $\ee_{2n-1}=0=1-\hgt(s)$ so that the bounds of Proposition \ref{bound} are tight.

The case $k=1$ in the next two results relates to   \cite[Theorem 2]{Nied86a} and \cite[Corollary 1]{Nied86b}.
\bp If $s\in D^n$ or 
$s:\N$ $\ra D$ 
and
$$\ee_n\leq \left\{\begin{array}{ll}
1-k &\mbox{ if } n \mbox{ is odd}\\
k &\mbox{ if } n \mbox{ is even}\\
\end{array} \right.$$
for some fixed integer $k\geq 1$ then $\hgt(s)=k$.
\ep
\begin{proof} The definition implies that $\hgt(s)\leq k$ and Proposition 
\ref{bound} gives $1-\hgt(s)\leq 1-k$.
\epr
\bp (Cf. \cite[Theorem 2]{Nied86a}, \cite[Corollary 1]{Nied86b}) If  $\lim _{n\ra\infty}\LC_n(s)=\infty$ and for all $n$, $\ee_n\geq 1-k$ for some fixed integer $k\geq 1$, then $\hgt(s)\leq k$.
\ep
\begin{proof}   Suppose that for some $n$, $\ee_n\geq k+1$.  Let $m\geq n$ be the first integer for which   $\Delta_{m+1}\neq 0$ (such an $m$ exists since $\lim_{n\ra\infty}\LC_n=\infty$). Then $\ee_m\geq\ee_n\geq  k+1$, so that $\ee_{m+1}=-\ee_m+1\leq -k$ which is a contradiction. Hence $\ee_n\leq k$ for all $n$ and $\hgt(s)\leq k$ as required.
\epr

\bp \label{last}(Cf. \cite[Theorem 3]{Nied86b}) 
 If $s\in D^n$ or $s:\N$ $\ra D$ is irrational then $s$ has a PLCP if and only if $\hgt(s)=1$.
\ep
\begin{proof} If $\hgt(s)=1$ then by Proposition \ref{bound}, $0\leq \ee_n\leq 1$ i.e. $\frac{n}{2}\leq \LC_n\leq\frac{n+1}{2}$ and 
$\LC_n=\lfloor \frac{n+1}{2}\rfloor$. The converse follows from Theorem \ref{basictfae}. 
\epr
\br We can now use  Propositions \ref{bound} --- \ref{last} and Theorem \ref{CFK} to deduce results of  \cite{Nied86a}, \cite{Nied86b}.
\er
\subsection{The LC Sum}
To simplify the notation in this section, we put $\LC_i=\LC_i(s)$, $\sigma_0=0$ and $$\sigma_n=\sigma_n(s)=\sum_{i=1}^{n}\LC_i$$
where $s\in D^n$. We will make repeated use of \cite[Theorem 2]{Ma69}: if $\Delta_n\neq 0$ then $\LC_n=\max\{\LC_{n-1},n-\LC_{n-1}\}$ or $\LC_n=\max\{e,0\}+\LC_{n-1}$ in the notation of Theorem \ref{bimrt}(i).
 We will need the fact that  $\sum_{i=1}^{n}\lfloor  \frac{i+1}{2}\rfloor= \lfloor \frac{(n+1)^2}{4}\rfloor$ which is easily proved by induction.
We begin with a technical lemma.
\begin{lemma}\label{ltwo}
 For integers $k\geq -1$ and $l\geq 1$,
$$\sum_{i=k+1}^{k+2l}\left\lfloor  \frac{i+1}{2}\right\rfloor=l^2+(k+1)l.$$
\end{lemma}
\bpr
If we put $m=k+l+1$, the sum is
\begin{eqnarray*}\label{two}
\sum_{i=0}^{l-1}\left( \left\lfloor \frac{m-i}{2}\right\rfloor + 
			\left\lfloor \frac{m+i+1}{2}\right\rfloor\right)=			\sum_{i=0}^{l-1}\left( \frac{m-i}{2} + 
			\frac{m+i+1}{2}-\frac{1}{2}\right)=km
\end{eqnarray*}
since $m-i$ and $m+i+1$ have opposite parity. 
\epr

 \bt\label{stable} For $s\in D^n$, 
$\sigma_n\leq \sum_{i=1}^{n}\lfloor  \frac{i+1}{2}\rfloor$.
\et
\bpr It is convenient to set $\sigma_{-1}=\LC_{-1}=0$. Let us call $j\geq -1$ {\em stable}  if it is odd, $\LC_j=\lfloor   \frac{j+1}{2}\rfloor$ and $\sigma_j \leq \sum_{i=1}^{j}\lfloor   \frac{i+1}{2}\rfloor$.
Clearly $-1$ is stable. Suppose inductively
that $k=2c-1\geq -1$ is stable, so that $\LC_k=c$. We show that (i) we can assume that there is an $l\geq 2$ such that $\LC_i=c$ for $k+1\leq i\leq k+l<n$, $\LC_{k+l+1}\neq c$ and either (ii)  $k+2l\leq n$ and $k+2l$ is stable or (iii)  $n<k+2l$ and $n$ is stable. This will complete the proof as we can replace $k$ by $k+2l$ and $c$ by $c+l$ in (i) until for some $k$ and $l$, $n<k+2l$. Then (iii) applies and we conclude that $n$ is stable. 

(i) By Theorem \ref{bimrt}(i), $\LC_{k+1}=c$ independently of $\Delta_{k+1}$. We are done if $n=k+1$, so suppose that $n\geq k+2$. If $\Delta_{k+2}\neq 0$ then $\LC_{k+2}=c+1=\lfloor  \frac{k+2+1}{2}\rfloor$ and we can replace $k$ by $k+2$ and $c$ by $c+1$. So we can
assume that $\Delta_{k+2}=0$. Hence for some $l\geq 2$,
$\LC_i=c$ for $k\leq i\leq k+l$. If $k+l=n$, then $\sigma_n\leq \sum_{i=1}^{n}\lfloor  \frac{i+1}{2}\rfloor$ since $k$ is stable. So we can assume that $k+l<n$ and $\LC_{k+l+1}\neq c$. 

 (ii) Let $n\geq k+2l$.  Firstly,  $\Delta_{k+l+1}\neq 0$ since $\LC_{k+l+1}\neq c$. Hence $\LC_{k+l+1}=c+l$ by  Theorem \ref{bimrt}(i). As $c+l=\max\{c+l,k+2l-(c+l)\}$, we have  
\begin{eqnarray}\label{Lsum}\LC_i=\left\{\begin{array}{ll}
c & \mbox{ for } k\leq i\leq k+l\\
c+l& \mbox{ for } k+l< i\leq k+2l
\end{array}\right.
\end{eqnarray}
and in particular, $\LC_{k+2l}=\lfloor \frac{k+2l+1}{2}\rfloor$.  Equation (\ref{Lsum}) and Lemma \ref{ltwo} imply that
$$\sum_{i=k+1}^{k+2l}\LC_i=lc+l(c+l)=l^2+(k+1)l=\sum_{i=k+1}^{k+2l}\left\lfloor  \frac{i+1}{2}\right\rfloor$$
and since $k$ is stable, 
$\sigma_{k+2l}=\sigma_k+l^2+(k+1)l\ \leq\   \sum_{i=1}^{k+2l}\left\lfloor   \frac{i+1}{2}\right\rfloor$
i.e. $k+2l$ is stable. 

(iii) Suppose now that $n<k+2l$ and let $m=n-k-l-1$, so that $0\leq m< l-1$. As $k$ is stable, it is enough to show that
$\sum_{i=k+1}^{n}\LC_i
 \leq \sum_{i=k+1}^{n}\lfloor  \frac{i+1}{2}\rfloor$. But
 $$\sum_{i=k+1}^{n}\LC_i=\sum_{i=k+1}^{k+l-m-1}\LC_i+\sum_{i=0}^{m}\left(\LC_{k+l-i}+\LC_{k+l+i+1}\right)$$
and for $k+1\leq i\leq k+l-m-1$, $\LC_i=c\leq \lfloor\frac{i+1}{2}\rfloor$. Further, Equation (\ref{Lsum}) implies that the second summand is $(m+1)(2c+l)=(m+1)(k+l+1)$. As in the proof of Lemma \ref{ltwo}, 
$$\sum_{i=0}^{m}\left(\left\lfloor\frac{k+l-i+1}{2}\right\rfloor+\left\lfloor\frac{k+l+i+2}{2}\right\rfloor\right)=(m+1)(k+l+1).$$ Therefore $\sum_{i=k+1}^{n}\LC_i
 \leq \sum_{i=k+1}^{n}\lfloor  \frac{i+1}{2}\rfloor$ and the proof is complete.
\epr

The following consequence of Theorem \ref{stable} appeared in \cite{FJ98} for even $n$ and $D$ a field.
\begin{corollary}  \label{sum.l}For $s\in D^n$,  $\sum_{i=1}^{n}\LC_i \leq  \lfloor (n+1)^2/4\rfloor$.
\end{corollary}

Two more equivalent conditions for a PLCP follow.
\bc\label{char} Let $s\in D^n$. The following are equivalent:

(i) $s$ has a PLCP

(ii) $\LC_i\leq \lfloor  \frac{i+1}{2}\rfloor$ for $1\leq i\leq n$ and $\sigma_n= \lfloor \frac{(n+1)^2}{4}\rfloor$

(iii)  $\LC_i\geq \lfloor  \frac{i+1}{2}\rfloor$ for $1\leq i\leq n$.

\ec
\bpr Theorem \ref{stable} shows that (i) $\Rightarrow$ (ii). (ii) $\Rightarrow$ (i): The case $n=1$ is a trivial verification. Suppose inductively that the converse is true for sequences of length $n-1\geq 1$,    $\LC_i\leq \lfloor  \frac{i+1}{2}\rfloor$ for $1\leq i\leq n$ and $\sigma_n= \lfloor \frac{(n+1)^2}{4}\rfloor$.
Then $\lfloor \frac{n^2}{4}\rfloor+ \lfloor \frac{n+1}{2}\rfloor=\sigma_n=\sigma_{n-1}+\LC_n\leq  \lfloor\frac{n^2}{4}\rfloor+\LC_n$  by Theorem \ref{stable}. We conclude that $\LC_n= \lfloor \frac{n+1}{2}\rfloor$ and so $\sigma_{n-1}=\lfloor  \frac{n^2}{4}\rfloor$. The inductive hypothesis now shows that $s^{(n-1)}$ has a PLCP.

(i) $\Rightarrow$ (iii) is trivial. (iii) $\Rightarrow$ (i): This is trivial for $n=1$. Suppose inductively that the converse is true for sequences of length $n-1\geq 1$ and  that  $\LC_i\geq\lfloor  \frac{i+1}{2}\rfloor$ for $1\leq i\leq n$.
Then $\sigma_n\geq\sum_{i=1}^n\lfloor \frac{n+1}{2}\rfloor= \lfloor \frac{(n+1)^2}{4}\rfloor$ and therefore $\sigma_n= \lfloor \frac{(n+1)^2}{4}\rfloor$ by Theorem \ref{stable}. By the same argument, $\sigma_{n-1}= \lfloor \frac{n^2}{4}\rfloor$ and hence $\LC_n=\lfloor \frac{(n+1)^2}{4}\rfloor-\lfloor \frac{n^2}{4}\rfloor=\lfloor \frac{n+1}{2}\rfloor$. This together with the inductive hypothesis implies that $s$ has a PLCP.
\epr
\be Let $s$ be the geometric sequence $(1,1,1)$ with minimal polynomial $x+1$ and $\LC_1=\LC_2=\LC_3=1$. As $\LC_3\neq 2$, $s$ does not have a PLCP.  In fact both (i) $\sigma_3=\lfloor \frac{(3+1)^2}{4}\rfloor$ and (ii) $\LC_3\geq 2$ fail. Now let $t=(1,1,1,0)$ which does not have a PLCP since $s$ does not.  Algorithm MP gives $x^3+x^2+1\in \Min(t)$, so $\LC_4=3$. We have $\sigma_4=6=\lfloor  \frac{(4+1)^2}{4}\rfloor$, but both (i) $\LC_4\leq 2$ and (ii) $\LC_3\geq 2$ fail.
\eex
\begin{remark}
 Let $D$ be a field $\F$. Corollary \ref{sum.l} easily gives an upper bound for the number of $\F$-multiplications required by Algorithm MP on a sequence of length $n$. We can clearly  divide by $\Delta'_{i+1}$ at iteration $i$, which requires at most $\LC_i+1$ $\F$-multiplications to compute the discrepancy and at most a further $\LC_{i'}+1\leq \LC_i$ for the updating when $\ee_i>0$.  Ignoring subquadratic terms, this gives a worst-case upper bound of $\lceil \frac{n^2}{2}\rceil$ $\F$-multiplications. 
 \er

\end{document}